\documentclass[review]{elsarticle}

\usepackage{hyperref}
\usepackage{lineno}
\modulolinenumbers[5]

\usepackage{graphicx}
\usepackage{bm}
\hypersetup{colorlinks=false}

\journal{General Relativity and Gravitation}

\begin{document}

\begin{frontmatter}

\title{Resonant frequencies of a massless scalar field in the canonical acoustic black hole spacetime}


\author[mymainaddress]{H. S. Vieira\corref{mycorrespondingauthor}}
\cortext[mycorrespondingauthor]{Corresponding author}
\ead{horacio.santana.vieira@hotmail.com}

%\author[mymainaddress]{D. A. T. Vanzella}
%\ead{vanzella@ifsc.usp.br}

\author[secondaryaddress]{V. B. Bezerra}
\ead{valdir@fisica.ufpb.br}

\address[mymainaddress]{Instituto de F\'{i}sica de S\~{a}o Carlos, Universidade de S\~{a}o Paulo, Caixa Postal 369, CEP 13560-970, S\~{a}o Carlos, S\~{a}o Paulo, Brazil}
\address[secondaryaddress]{Departamento de F\'{i}sica, Universidade Federal da Para\'{i}ba, Caixa Postal 5008, CEP 58051-970, Jo\~{a}o Pessoa, PB, Brazil}

\begin{abstract}
In this work we consider the exact solution of the Klein-Gordon equation describing a massless scalar field in the spacetime of a four dimensional canonical acoustic black hole, which is given in terms of the general Heun function, to investigate the interesting phenomena related to the resonant frequencies.
\end{abstract}


\end{frontmatter}

%\linenumbers

%
%%%%%%%%%%%%%%%%%%%%%%%%%%%%%%%%%%%%%%%%%%%%%%%%%%%%%%%%%%%%%%%%%%%%%%%%%%%%%%%%%%%%%%%%%%%%%% Introduction
%
\section{Introduction}
Even with the recent detection of astrophysical black holes \cite{PhysRevLett.116.221101,AstrophysJLett.875.L1}, it is still extremely difficult to measure some physical phenomena which arise from the interaction between quantum fields and gravitational spacetime backgrounds like compact objects and black holes. In order to overcome these issues, W. G. Unruh \cite{PhysRevLett.46.1351} proposed some analogue models of gravity, where the equation of motion which describes the propagation of sound modes (phonons) on a background hydrodynamic flow (acoustic black hole) can be written as formally identical to the massless Klein-Gordon equation, whose field propagates in a (3+1)-dimensional Lorentzian geometry. In this context, in which case gravity can emerge as a phenomenon associated to somed excitations, it would be possible, in principle, to observe in the laboratory some effects which mimic the know phenomena of quantum gravity \cite{ClassQuantGrav.16.3953,JHEP.1006.087,NewJPhys.12.095014,NewJPhys.12.095012,PhysLettB.694.149,PhysRevD.85.025013,ChinPhysC.41.043105,IntJModPhysD.26.1750035}. 

Since the seminal paper by W. G. Unruh, which opened up this field of research, a lot of investigations in which concern the physics of acoustic black holes was performed, as for example, the Hawking-Unruh radiation from a rotating acoustic black hole \cite{PhysLettB.698.438}, the relativistic acoustic metric for a planar black hole \cite{PhysLettB.752.13}, the quantum potential in analogue gravity \cite{PhysRevD.96.064027}, the canonical acoustic thin-shell wormholes \cite{ModPhysLettA.32.1750047}, and the holographic transport \cite{PhysRevD.100.056015}. Among these studies, two are of special interest for us, namely, the work where analytic solutions for massless scalar fields in the both canonical and rotating acoustic black holes are obtained in terms of the Heun's functions \cite{GenRelGrav.48.88,Correction}, and the one about probing the Unruh effect with an accelerated extended system \cite{NatureComm.10.3030}.

It is worth point out one important aspect, which comes out from the gravity analogue phenomena associated with these analogue models, that is related to the possibility of using these analogies to try probe and understand some features of quantum field theory in curved spacetime by the realization of accurate experiments. As an example of such a system, which can be used with this proposal, we can mention the Bose-Einstein Condensate \cite{c}. Other condensed matter systems, such as super-fluid helium \cite{d,e} and degenerate Fermi gas \cite{f}, have also been used with this proposal.

Inspired by all of these previous works and adopting our technique developed in \cite{AnnPhys.373.28}, we will focus on the computation of the resonant frequencies for a massless scalar field in the spacetime of a canonical acoustic black hole. In this scenario there are two parameters: the radius of the acoustic event horizon and the speed of sound. It is worth emphasizing that these two parameters are fully controllable in the laboratory, and hence can be measured.

The dynamics arising from the interaction of fundamental quantum fields and black holes can be partially understood by examining the response of the black hole to external perturbations, given in terms of some characteristic oscillations, the so called quasinormal modes (QNMs) to which we can associate the corresponding resonant frequencies, for the case where the oscillations decay exponentially \cite{PhysRevD.94.084040}, characterizing the linearized relaxation dynamics \cite{1,2}. These oscillations are expected to appear in different astrophysical process produced by black holes, and undoubtedly, they are of great importance not only from the astrophysical point of view themselves \cite{3,4}, but also from the theoretical as well \cite{5,6}.

The study of the QNMs of black holes began in the 1970's \cite{7,8}. From these days until now, the role played by these characteristics oscillations confirmed that they are an essential tool to understand the dynamics of astrophysical black holes. In particular, the gravitational emission of radiation by a collision of black holes can, certainly, be better understood if we consider the resonant frequencies which dominates this phenomenon. Also, these QNMs could provide some information about the physical parameters which identify, uniquely, the black hole, such as mass, charge and angular momentum.

Thus, due to the important role played by the QNMs, this topic has received a great attention in the last years \cite{9,14,10} and the computation of quasinormal frequencies was performed \cite{15,16}.

These characteristic oscillations also play a role in the dynamics of analogue black holes. Studies in these backgrounds include the investigation of quasinormal modes \cite{17}, relation between these modes and the angular momentum \cite{18} and about the connections between these modes and the Regge poles \cite{PhysRevD.82.084037}. In all three cases by considering a canonical acoustic black hole.

In order to obtain these resonant frequencies, we need to impose that the radial solution of the Klein-Gordon equation, which is given in terms of the general Heun function \cite{Ronveaux:1995}, should have a polynomial form. These special functions of mathematical physics have gained increasingly more importance due to their large number of applications in different areas of natural science, from biology to physics \cite{AdvHighEnergyPhys.2018.8621573}.

In this work we will consider a stationary and asymptotically flat canonical acoustic black hole \cite{20}. It comes out if we consider a spherically symmetric steady flow of incompressible fluid in three dimensions, with a source at $r=0$.

The paper is organized as follows. In Section \ref{Scalar_solution} we present the canonical acoustic metric and the analytical solution of the massless Klein-Gordon equation. In Section \ref{Resonant_frequencies} we compute the resonant frequencies (damped quasinormal frequencies). In Section \ref{Conclusions} we present a brief discussion. Here we adopt the natural units where $G=c=\hbar=1$.
%
%%%%%%%%%%%%%%%%%%%%%%%%%%%%%%%%%%%%%%%%%%%%%%%%%%%%%%%%%%%%%%%%%%%%%%%%%%%%%%%%%%%%%%%%%%%%%% Scalar solution in the four dimensional canonical acoustic black hole
%
\section{Scalar solution in the four dimensional canonical acoustic black hole}\label{Scalar_solution}
Recently, we have studied the behavior of a massless scalar field in the canonical acoustic black hole spacetime \cite{GenRelGrav.48.88}. For this background, the line element which appropriately describes the solution for a spherically symmetric flow is given by
\begin{equation}
ds^{2} = -c^{2}\Delta\ d\tau^{2}+\Delta^{-1}\ dr^{2}+ r^{2}d\theta^{2}+r^{2}\sin^{2}\theta\ d\phi^{2},
\label{eq:metrica_canonical_Schw}
\end{equation}
with
\begin{equation}
\Delta=1-\frac{r_{0}^{4}}{r^{4}},
\label{eq:Delta_canonical_Schw}
\end{equation}
where
\begin{equation}
r_{h}=r_{0}
\label{eq:horizon_canonical_Schw}
\end{equation}
is the radius of the acoustic event horizon. Here $c$ is the speed of sound, which is constant throughout the fluid flow.

Note that this form, which is similar to Schwarzschild metric, is obtained by appropriately choosing a time coordinate \cite{20}. It is worth commenting that this fourth power comes out from the fluid velocity ($v=cr_{0}^{2}/r^{2}$) and turns this metric completely different from Schwarzschild one.

For a massless scalar field, the Klein-Gordon equation can be write as
\begin{equation}
\biggl[\frac{1}{\sqrt{-g}}\partial_{\rho}(g^{\rho\sigma}\sqrt{-g}\partial_{\sigma})\biggr]\Psi=0.
\label{eq:Klein-Gordon_cova_draining_bathtub_Kerr}
\end{equation}
The spacetime is static and symmetric under time translations, and thus the solution corresponding to the temporal part is given by $\mbox{e}^{-i \omega \tau}$, where $\omega$ is the frequency. In fact, $\omega$ is the energy because we are considering $\hbar=1$. Additionally, the spacetime given by metric (\ref{eq:metrica_canonical_Schw}) also has rotational symmetry with respect to the azimuthal angle, $\phi$, and thus the solution is $\mbox{e}^{i \omega \phi}$, where $m=\{\pm 1,\pm 2,\ldots\}$ is the azimuthal quantum number. Thus, the general angular solution is given in terms of the spherical harmonic function $Y_{m}^{l}(\theta,\phi)=P_{m}^{l}(\cos\theta)\mbox{e}^{im\phi}$, where $l$ is an integer such that $|m| \leq l$. Therefore, the scalar wave function can be written as
\begin{equation}
\Psi(\mathbf{r},\tau)=R(r)Y_{m}^{l}(\theta,\phi)\mbox{e}^{-i \omega \tau}.
\label{eq:separation}
\end{equation}
Then, substituting Eq.~(\ref{eq:separation}) into Eq.~(\ref{eq:Klein-Gordon_cova_draining_bathtub_Kerr}), we obtain
\begin{equation}
\frac{d}{dr}\biggl(r^{2}\Delta\frac{dR}{dr}\biggr)+\biggl(\frac{\omega^{2}r^{2}}{c^{2}\Delta}-\lambda_{lm}\biggr)R=0.
\label{eq:mov_radial_1_canonical_Schw}
\end{equation}

Now, we define a new radial coordinate, $z$, by setting the following homographic substitution
\begin{equation}
z=\frac{x-a_{1}}{a_{2}-a_{1}}=\frac{r^{2}-r_{0}^{2}}{-2r_{0}^{2}},
\label{eq:z_canonical_Schw}
\end{equation}
where $x=r^{2}$ and $(a_{1},a_{2})=(x_{+},x_{-})$, with $x_{\pm}=\pm r_{0}^{2}$. Furthermore, we perform a transformation of the dependent variable, $R(z)$, given by
\begin{equation}
R(z)=z^{-\frac{1}{2}}(z-1)^{-\frac{1}{2}}(z-a)^{-\frac{3}{4}}U(z).
\label{eq:F-homotopic_canonical_Schw}
\end{equation}
Thus, we obtain the following equation for the function $U(z)$
\begin{eqnarray}
&& \frac{d^{2}U}{dz^{2}}+\biggl\{\frac{(1+4B_{1}^{2})/4}{z^{2}}+\frac{(1+4B_{2}^{2})/4}{(z-1)^{2}}\nonumber\\
&& +\frac{(3+16B_{3}^{2})/16}{(z-a)^{2}}\nonumber\\
&& +\frac{(A_{1}+A_{3})r_{0}^{2}+a[-2+(A_{1}+A_{2})r_{0}^{2}]}{a(z-1)(z-a)}\nonumber\\
&& +\frac{3+2a-4A_{1}r_{0}^{2}}{4z(z-1)(z-a)}\biggr\}R=0,
\label{eq:mov_radial_x_canonical_Schw_4}
\end{eqnarray}
where $a=1/2$. The coefficients $A_{1}$, $A_{2}$, $A_{3}$, $B_{1}$, $B_{2}$, and $B_{3}$ are expressed as
\begin{equation}
A_{1}=\frac{-2c^{2} \lambda_{lm} r_{0}^{2}-2 c^{2} \lambda_{lm}  r_{0}^{2}-4 \omega ^{2}}{32 c^{2} r_{0}^{8}},
\label{eq:A1_canonical_Schw_3}
\end{equation}
\begin{equation}
A_{2}=\frac{2c^{2} \lambda_{lm}  r_{0}^{2}+2 c^{2} \lambda_{lm}  r_{0}^{2}-4 \omega ^{2}}{32 c^{2} r_{0}^{8}},
\label{eq:A2_canonical_Schw_3}
\end{equation}
\begin{equation}
A_{3}=\frac{\omega ^{2}}{4 c^{2} r_{0}^{8}},
\label{eq:A3_canonical_Schw_3}
\end{equation}
\begin{equation}
B_{1}=\frac{\omega}{4 c r_{0}^{3}},
\label{eq:B1_canonical_Schw_3}
\end{equation}
\begin{equation}
B_{2}=\frac{\omega}{4 i c r_{0}^{3}},
\label{eq:B2_canonical_Schw_3}
\end{equation}
\begin{equation}
B_{3}=-\frac{1}{2 r_{0}^{2}}\sqrt{\lambda_{lm}},
\label{eq:B3_canonical_Schw_3}
\end{equation}
with $\lambda_{lm}=l(l+1)$.

Equation (\ref{eq:mov_radial_x_canonical_Schw_4}) is a general Heun equation \cite{Ronveaux:1995}. Thus, the radial solution of the massless Klein-Gordon equation in the spacetime corresponding to the canonical acoustic black hole is given by \cite{Correction}
\begin{eqnarray}
R(z) & = & z^{\frac{1}{2}(\gamma-1)}(z-1)^{\frac{1}{2}(\delta-1)}(z-a)^{\frac{1}{2}(\epsilon-2)}\nonumber\\
& & \times \{C_{1}\ \mbox{HeunG}(a,q;\alpha,\beta,\gamma,\delta;z)\nonumber\\
& & + C_{2}\ z^{1-\gamma}\ \mbox{HeunG}(a,q_{1};\alpha_{1},\beta_{1},\gamma_{1},\delta;z)\},\nonumber\\
\label{eq:solucao_geral_radial_canonical_Schw}
\end{eqnarray}
where $\mbox{HeunG}(a,q;\alpha,\beta,\gamma,\delta;z)$ is the general Heun function, $C_{1}$ and $C_{2}$ are constants. The parameters $\alpha$, $\beta$, $\gamma$, $\delta$, $\epsilon$, and $q$ are given by
\begin{eqnarray}
\alpha & = & \frac{1}{2}\biggl\{\biggl[\frac{25}{4}-4( A_{1} r_{0}^{2}+ A_{2} r_{0}^{2}+B_{1}^2+ B_{2}^2\nonumber\\
&& +B_{3}^2)-8 r_{0}^{2} (A_{1} + A_{3})\biggr]^{\frac{1}{2}}\nonumber\\
&& +\frac{1}{2} [4 i (B_{1}+ B_{2})+ 4+\sqrt{1-16 B_{3}^2}]\biggr\},
\label{eq:alpha}
\end{eqnarray}
\begin{equation}
\beta=\frac{1}{2} [4 i (B_{1}+ B_{2})+\sqrt{1-16 B_{3}^2}+4]-\alpha,
\label{eq:beta_radial_HeunG_canonical_Schw}
\end{equation}
\begin{equation}
\gamma=1+2iB_{1},
\label{eq:gamma_radial_HeunG_canonical_Schw}
\end{equation}
\begin{equation}
\delta=1+2iB_{2},
\label{eq:delta_radial_HeunG_canonical_Schw}
\end{equation}
\begin{equation}
\epsilon=1+\frac{1}{2}\sqrt{1-16B_{3}^{2}},
\label{eq:epsilon_radial_HeunG_canonical_Schw}
\end{equation}
\begin{eqnarray}
q & = & \frac{1}{4} \biggl\{2 i B_{1} \biggl[\frac{1}{2} (2+4 i B_{2})+\sqrt{1-16 B_{3}^2}+2\biggr]\nonumber\\
&& +4 i \frac{1}{2} B_{2}+4 A_{1} r_{0}^{2}+\sqrt{1-16 B_{3}^2}-1\biggr\}.
\label{eq:q_radial_HeunG_canonical_Schw}
\end{eqnarray}
Furthermore, the parameters $\alpha_{1}$, $\beta_{1}$, $\gamma_{1}$, and $q_{1}$ are given by the relations
\begin{equation}
\alpha_{1}=\alpha+1-\gamma,
\label{eq:alpha_1_general_Heun}
\end{equation}
\begin{equation}
\beta_{1}=\beta+1-\gamma,
\label{eq:beta_1_general_Heun}
\end{equation}
\begin{equation}
\gamma_{1}=2-\gamma,
\label{eq:gamma_1_general_Heun}
\end{equation}
\begin{equation}
q_{1}=q+(\alpha\delta+\epsilon)(1-\gamma).
\label{eq:q_1_general_Heun}
\end{equation}

In what follows, we will use this radial solution and our recently developed technique \cite{AnnPhys.373.28} to compute the resonant frequencies of massless scalar particles propagating in a canonical acoustic black hole.
%
%%%%%%%%%%%%%%%%%%%%%%%%%%%%%%%%%%%%%%%%%%%%%%%%%%%%%%%%%%%%%%%%%%%%%%%%%%%%%%%%%%%%%%%%%%%%%% Resonant frequencies
%
\section{Resonant frequencies}\label{Resonant_frequencies}
The resonant frequencies are determined from the radial solution, given by Eq.~(\ref{eq:solucao_geral_radial_canonical_Schw}), subject to two boundary conditions, namely, that it should be finite in the region exterior to the event horizon and well behaved at infinity, that is, $R(z) < \infty$ for $|z| > 0$ as well as for $|z| \rightarrow \infty$.

Let us examine these conditions. If $\gamma \neq \{0,-1,-2,\ldots\}$, then from the Fuchs-Frobenius theory, it follows that $\mbox{HeunG}(a,q;\alpha,\beta,\gamma,\delta;z)$ exists, is analytic in the disk $|z| < 1$, corresponds to exponent $0$ at $z=0$, assumes the value $1$ there, and has the Maclaurin expansion
\begin{equation}
\mbox{HeunG}(a,q;\alpha,\beta,\gamma,\delta;z)=\sum_{j=0}^{\infty}b_{j}z^{j},
\label{eq:serie_HeunG_todo_x}
\end{equation}
where $b_{0}=1$, and
\begin{eqnarray}
	a\gamma b_{1}-qb_{0}=0,\nonumber\\
	X_{j}b_{j+1}-(Q_{j}+q)b_{j}+P_{j}b_{j-1}=0, \quad j \geq 1,
\label{eq:recursion_General_Heun}
\end{eqnarray}
with
\begin{eqnarray}
	P_{j}=(j-1+\alpha)(j-1+\beta),\nonumber\\
	Q_{j}=j[(j-1+\gamma)(1+a)+a\delta+\epsilon],\nonumber\\
	X_{j}=a(j+1)(j+\gamma)\ .
\label{eq:P_Q_X_recursion_General_Heun}
\end{eqnarray}
Thus, from this expansion, we conclude that the radial solution is finite when $r \rightarrow r_{h}$, which implies that $z \rightarrow 0$. Therefore, the first condition is satisfied. In fact, this condition yield the ingoing wave solution of the Hawking radiation, as was shown recently \cite{GenRelGrav.48.88}.

On the other hand, the second condition requires that $R(z)$ must have a polynomial form. Indeed, the general Heun function, $\mbox{HeunG}(a,q;\alpha,\beta,\gamma,\delta;z)$, becomes a polynomial of degree $n$ if the following condition is satisfied:
\begin{equation}
\alpha=-n,
\label{eq:condiction_poly_General_Heun}
\end{equation}
with $n=\{0,1,2,\ldots\}$.

Therefore, substituting Eq.~(\ref{eq:alpha}) into Eq.~(\ref{eq:condiction_poly_General_Heun}), we can find an expression for the resonant frequencies associated to massless scalar particles in the four dimensional canonical acoustic black hole. It is given by 
\begin{equation}
\omega_{n} = \biggl(\frac{1}{2}+\frac{i}{2}\biggr) c r_{0} [\sqrt{r_{0}^4-4 l(l+1)} + (4 n +9) r_{0}^2],
\label{eq:RFs_canonical_Schw}
\end{equation}
where $n$ is the principal quantum number.

This frequency (energy) spectrum gives a complex number, such that $\omega=\omega_{R}+i\ \omega_{I}$, where $\omega_{R}$ and $\omega_{I}$ are the real and imaginary parts, respectively. We remark that the eigenvalues given by Eq.~(\ref{eq:RFs_canonical_Schw}) are degenerate, since that there is a dependence on the eigenvalue $\lambda_{lm}$.

These results are new, and hence we do not have any similar one obtained in the literature to compare. It is worth emphasizing that, in this work, we obtain the resonant frequencies directly from the general Heun function by using the condition which should be imposed in such a way that this function reduces to a polynomial. In addition, our results are complete, in the sense that they include a discussion about the dependence of the resonant frequencies on the radius of the acoustic event horizon and the speed of sound.

The resonant frequencies for $n=0$ and $n=7$ are shown in Tables \ref{tab:Rfs_canonical_Schw_1} and \ref{tab:Rfs_canonical_Schw_2}, respectively.

\begin{table}[!ht]
\caption{Values of the massless scalar resonant frequencies $\omega_{0}$ for $l=\{0,1,2,3,4,5,6\}$. The units are in multiples of $c=r_{0}=1$.}
\label{tab:Rfs_canonical_Schw_1}
\begin{tabular}{ccc}
\hline\noalign{\smallskip}
			$l$  & $\mbox{Re}(\omega_{0})$ & $\mbox{Im}(\omega_{0})$ \\
\noalign{\smallskip}\hline\noalign{\smallskip}
			0 & 5.00000 & 5.00000 \\
			1 & 3.17712 & 5.82288 \\
			2 & 2.10208 & 6.89792 \\
			3 & 1.07217 & 7.92783 \\
			4 & 0.05590 & 8.94410 \\
			5 & -0.95436 & 9.95436 \\
			6 & -1.96142 & 10.96142 \\
\noalign{\smallskip}\hline
\end{tabular}
\end{table}

\begin{table}[!ht]
\caption{Values of the massless scalar resonant frequencies $\omega_{7}$ for $l=\{0,1,2,3,4,5,6\}$. The units are in multiples of $c=r_{0}=1$.}
\label{tab:Rfs_canonical_Schw_2}
\begin{tabular}{ccc}
\hline\noalign{\smallskip}
			$l$  & $\mbox{Re}(\omega_{7})$ & $\mbox{Im}(\omega_{7})$ \\
\noalign{\smallskip}\hline\noalign{\smallskip}
			0 & 19.00000 & 19.00000 \\
			1 & 17.17712 & 19.82288 \\
			2 & 16.10208 & 20.89792 \\
			3 & 15.07217 & 21.92783 \\
			4 & 14.05590 & 22.94410 \\
			5 & 13.04564 & 23.95436 \\
			6 & 12.03858 & 24.96142 \\
\noalign{\smallskip}\hline
\end{tabular}
\end{table}

From Tables \ref{tab:Rfs_canonical_Schw_1} and \ref{tab:Rfs_canonical_Schw_2} we see that the real part of the resonant frequencies decreases with $l$, while the imaginary part increases very quickly for fixed values of the radius of the acoustic event horizon and the speed of sound. Therefore, we conclude that the canonical acoustic black hole is stable for some values of the eigenvalue $l$, that is, when $r_{0}^4 \geq 4 l(l+1)$. Otherwise, it shows the instability of this canonical acoustic black hole.

\begin{figure}[!ht]
	\includegraphics[scale=0.9]{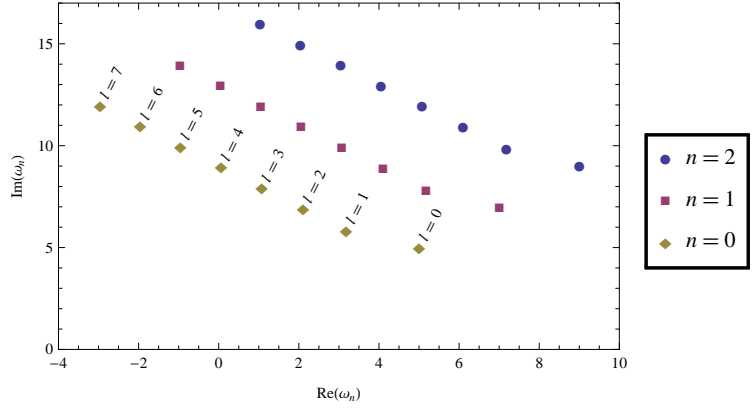}
	\caption{The massless scalar resonant frequencies as a function of $l$ for $n=\{0,1,2\}$. The units are in multiples of $c=r_{0}=1$.}
	\label{fig:Fig1_Acoustic_hole_II}
\end{figure}

\begin{figure}[!ht]
	\includegraphics[scale=0.9]{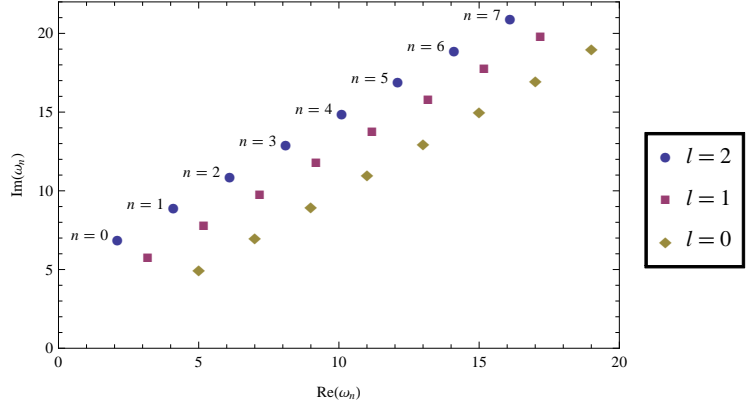}
	\caption{The massless scalar resonant frequencies as a function of $n$ for $l=\{0,1,2\}$. The units are in multiples of $c=r_{0}=1$.}
	\label{fig:Fig2_Acoustic_hole_II}
\end{figure}

The resonant frequencies that we have obtained are presented in Figures (\ref{fig:Fig1_Acoustic_hole_II}) and (\ref{fig:Fig2_Acoustic_hole_II}), as a function of $n$ and $l$, respectively.

%\newpage

In Figures \ref{fig:Fig1_Acoustic_hole_II} and \ref{fig:Fig2_Acoustic_hole_II} we see that for fixed values of the radius of the acoustic event horizon and the speed of sound, the real part decreases with $l$, while the imaginary part increases thus the decay rate \cite{PhysLettB.608.10} of the field is growing.
%
%%%%%%%%%%%%%%%%%%%%%%%%%%%%%%%%%%%%%%%%%%%%%%%%%%%%%%%%%%%%%%%%%%%%%%%%%%%%%%%%%%%%%%%%%%%%%% Conclusions
%
\section{Conclusions}\label{Conclusions}
We have shown that the four dimensional canonical acoustic black hole presents an oscillating energy spectrum associated to the QNMs with complex frequencies corresponding to damped oscillations, the resonant frequencies.

This wave phenomena is due to the interaction between massless scalar fields and the effective geometry of the background under consideration. The obtained resonant frequencies are degenerate, and depend on the radius of the acoustic event horizon and speed of sound.

Recently \cite{NaturePhys.10.864,PhysRevD.92.024043,Nature.569.688} it was observed the thermal Hawking-Unruh radiation and its temperature in an analogue black hole. Therefore, we believe that it will be possible, in a near future, to measure these resonant frequencies in laboratory.
%
%%%%%%%%%%%%%%%%%%%%%%%%%%%%%%%%%%%%%%%%%%%%%%%%%%%%%%%%%%%%%%%%%%%%%%%%%%%%%%%%%%%%%%%%%%%%%% acknowledgments
%
\section*{Acknowledgement}
H.S.V. is funded by the Coordena\c c\~{a}o de A\-per\-fei\-\c co\-a\-men\-to de Pessoal de N\'{i}vel Superior - Brasil (CAPES) - Finance Code 001. V.B.B. is partially supported from CNPq Project No. 305835/2016-5. The authors also would like to thank Professor Daniel Vanzella for the very fruitful discussions.
%
%%%%%%%%%%%%%%%%%%%%%%%%%%%%%%%%%%%%%%%%%%%%%%%%%%%%%%%%%%%%%%%%%%%%%%%%%%%%%%%%%%%%%%%%%%%%%% thebibliography
%

%
%%%%%%%%%%%%%%%%%%%%%%%%%%%%%%%%%%%%%%%%%%%%%%%%%%%%%%%%%%%%%%%%%%%%%%%%%%%%%%%%%%%%%%%%%%%%%%
%
\end{document}